\input harvmac
\input epsf

\def\p{\partial}

\def\half{{1\over 2}}

\def\({\left(}
\def\){\right)}
%%%%%%%%%%%%%%%%%%%%%%%%%%%%%%%%%%%%%%%%%%%%%%%%%%%%%%%%%%%%%%%%%%%%%

\Title{}{\vbox{\centerline{Accelerating universe emergent from the
landscape}}}

\centerline{ Jian-Huang She}
\bigskip \centerline{\it Institute of Theoretical Physics,
Academia Sinica} \centerline{\it P. O. Box 2735, Beijing 100080}

\medskip

 \centerline{\tt
jhshe@itp.ac.cn}

\bigskip

We propose that the existence of the string landscape suggests the
universe can be in a quantum glass state, where an extremely large
viscosity is generated, and long distance dynamics slows down. At
the same time, the short distance dynamics is not altered due to the
separation of time scales. This scenario can help to understand some
controversies in cosmology, for example the natural existence of
slow roll inflation and dark energy in the landscape, the apparent
smallness of the cosmological constant. We see also that moduli
stabilization is no longer necessary. We further identify the glass
transition point, where the viscosity diverges, as the location of
the cosmic horizon. We try to reconstruct the geometry of the
accelerating universe from the structure of the landscape, and find
that the metric should have an infinite jump when crossing the
horizon. We predict that the static coordinate metric for dS space
breaks down outside the horizon.

\Date{Jan, 2007}

%\draft

\nref\bp{R. Bousso, J. Polchinski, JHEP 0006:006,2000,
hep-th/0004134.}

\nref\kklt{S. Kachru, R. Kallosh, A. Linde and S. Trivedi,
Phys.Rev.D 68(2003)046005, hep-th/0301240. }

\nref\suss{L. Susskind, hep-th/0302219. }

\nref\dk{M. R. Douglas, S. Kachru, hep-th/0610102.}

\nref\tyeqc{H. Firouzjahi, S. Sarangi, S.-H. H. Tye, JHEP 0409
(2004) 060, hep-th/0406107; S. Sarangi, S.-H. H. Tye,
hep-th/0603237.}

\nref\fb{O. Lunin, S. D. Mathur, Nucl.Phys. B623 (2002) 342-394,
hep-th/0109154.}

\nref\bala{V. Balasubramanian, J. de Boer, V. Jejjala, J. Simon,
JHEP 0512:006,2005, hep-th/0508023; P. J. Silva, JHEP 0511:012,2005,
hep-th/0508081; P. G. Shepard, JHEP 0510:072,2005, hep-th/0507260.}

\nref\llm{H. Lin, O. Lunin, J. M. Maldacena, JHEP 0410:025,2004,
hep-th/0409174.}

\nref\gol{M. Goldstein, J. Chem. Phys. 51 3728(1969).}

\nref\wein{Steven Weinberg, ``Gravitation and Cosmology: Principles
and Applications of the General Theory of Relativity'', 1972.}

\nref\sti{F. H. Stillinger, T. A. Weber, Phys. Rev. A 25,
978(1982).}

\nref\kt{B. Freivogel, L. Susskind, Phys.Rev. D70 (2004) 126007,
hep-th/0408133; S.-H. H. Tye, hep-th/0611148; G.L. Kane, M.J. Perry,
A.N. Zytkow, hep-th/0311152; G.L. Kane, M.J. Perry, A.N. Zytkow,
Phys.Lett. B609 (2005) 7-12, hep-ph/0408169.}

\nref\sn{Supernova Cosmology Project (S. Perlmutter et al.),
Astrophys.J. 483 (1997) 565, astro-ph/9608192; Supernova Search Team
(Adam G. Riess et al.), Astron.J.116:1009-1038,1998,
astro-ph/9805201. }

\nref\gkp{S. B. Giddings, S. Kachru, J. Polchinski, Phys.Rev. D66
(2002) 106006, hep-th/0105097.}

\nref\lin{A. Linde, hep-th/0503195.}

\nref\rs{L. Randall, R. Sundrum, Phys.Rev.Lett. 83 (1999) 3370-3373,
hep-ph/9905221; Phys.Rev.Lett. 83 (1999) 4690-4693, hep-th/9906064.}

\nref\verl{H. L. Verlinde, Nucl.Phys.B580:264-274,2000,
hep-th/9906182.}

\nref\kiefer{C. Kiefer, Class. Quant. Grav. 4 (1987) 1369; J. J.
Halliwell, Phys. Rev. D39 (1989) 2912.}

\nref\cdgkl{A. Ceresole, G. Dall'Agata, A. Giryavets, R. Kallosh, A.
Linde, Phys.Rev. D74 (2006) 086010, hep-th/0605266.}

\nref\kur{J. Kurchan, L. Laloux, cond-mat/9510079.}

\nref\cdl{S. R. Coleman, F. De Luccia, Phys.Rev.D21:3305,1980.}

\nref\hm{S.W. Hawking, I.G. Moss, Phys.Lett.B110:35,1982.}

\nref\bt{J.D. Brown, C. Teitelboim, Phys. Lett. B195:177-182,1987;
Nucl. Phys. B297:787-836,1988.}

\nref\ghtw{A. Gomberoff, M. Henneaux, C. Teitelboim, F. Wilczek,
Phys.Rev.D69:083520,2004, hep-th/0311011.}

\nref\gger{P. Goldbart, N. Goldenfeld, Phys. Rev. A 39, 1402(1989).}

\nref\zw{R. Zwanzig, Phys. Rev. 156, 190(1967).}

\nref\hst{A. Hosoya, M. Sakagami, M. Takao, Annals
Phys.154:229,1984.}

\nref\kurc{J. Kurchan, L. Laloux, cond-mat/9510079.}

\nref\rsz{G. Ruocco, F. Sciortino, F. Zamponi, C. De Michele, T.
Scopigno, J.Chem.Phys. 120, 10666 (2004), cond-mat/0401449.}

\nref\ag{G. Adam, and J. H. Gibbs, J. Chem. Phys. 43, 139 (1965).}

\nref\hliu{G. Festuccia, H. Liu, JHEP 0604 (2006) 044,
hep-th/0506202. }

\nref\suwi{L. Susskind, E. Witten, hep-th/9805114.}

\nref\kivel{D. Kivelson, S. A. Kivelson, X-L Zhao, Z. Nussimov, and
G. Tarjus, Physica A 219, 27 (1995).}

\newsec{Introduction}
Recently it was found that a large number of metastable dS spaces
can be constructed in string theory \bp\ \kklt. And subsequently the
universality and importance of this phenomenon was realized and the
concept of landscape was coined in \suss. A set of fields and a
potential was assumed to describe the string theory vacua, and the
space of these fields was named landscape \suss. Currently there is
still wide gap between landscape and the real world.

It is often assumed that one vacuum is selected from the landscape,
which then serves as the playground for the excitations. The vacuum
is characterized by the dimension of spacetime, a cosmological
constant, the matter content, the rank of the gauge group etc.. Then
the question comes about the selection rule of the vacuum.
Statistical surveys of the landscape were explored by some groups
(see \dk\ and references therein), where the number distribution of
vacua was discussed, and many observable effects, such as the scale
of SUSY breaking, the gauge group, the matter content, can be
estimated. Anthropic principle was invoked to explain why we are in
such a vacuum \suss, and quantum cosmology was also reformed to do
this \tyeqc.

 In this short
note we advocate an alternative view that some characteristics of
the vacuum of our real world, for example the smallness of the
cosmological constant, result from the collective dynamics of the
whole landscape. In black hole physics, the fuzzball idea was
proposed to explain the information puzzle \fb, where the apparent
unique geometry with a horizon was deconstructed as the
superposition of many smooth geometries with given global charge. In
the context of AdS/CFT, this idea was more explicitly realized
\bala\ due to the construction of all the half-BPS geometries \llm.
There the smooth geometries act as spacetime foam, and a general
probe is not sensitive to the details of the foam. A universal
description of the low energy physics is the effective singular
geometry. This idea actually provides a connection between the many
microscopic vacua and the apparent vacuum, replacing the vacuum
selection rules.

Here we generalize the idea to the nonsupersymmetric landscape and
propose that the vacuum of our universe with small positive
cosmological constant is an effective description of the low energy
physics of the landscape. The effects of the spacetime foam can be
capsulated by an effective viscosity, whose existence signals the
slowing down of the large distance dynamics. The vacuum is
effectively treated as a medium with internal structure and the
viscous flow is a cooperative behavior of the internal degrees of
freedom. So we are effectively introducing a hydrodynamic theory for
the vacuum. In fact at finite temperature in the quark gluon plasma
state, the large distance, long time dynamics of gauge theory is
also described by a hydrodynamic theory, where shear and bulk
viscosities are also used to characterize the transport properties
of the medium.

 In a
local rest frame, where the three-momentum density vanishes
$T_{0i}=0$, the stress tensor of a medium with diffusive viscosity
generally has the form
\eqn\stre{T_{ij}=\delta_{ij}p-\eta(\p_iu_j+\p_ju_i-
{2\over3}\delta_{ij}\p_ku_k)-\zeta\delta_{ij}\p_ku_k,}where $p$ is
the pressure, $u_i$ the flow velocity, and $\eta$ and $\zeta$ are
the shear and bulk viscosities. $T_{ij}$ is the spatial component of
the energy momentum tensor $T_{\mu\nu}$. The shear viscosity $\eta$
characterizes the diffusive relaxation of transverse momentum
density fluctuations, and the bulk viscosity $\zeta$ characterizes
the departure from equilibrium during a uniform expansion. The bulk
viscosity $\zeta$ vanishes for a scale invariant theory.

We draw analogy from the supercooled glass forming liquids, where a
potential energy landscape was also discovered \gol\ \sti. The
defining property of the glass transition is the slowing down of the
dynamics. On cooling, the characteristic time scale, eg. relaxation
time and of the viscosity, increases by more than 15 orders of
magnitude in a relatively small temperature interval. We note that
in the cosmological setup, thermal fluctuations are replaced by
quantum fluctuations.

The layout of the note is as follows. In section 2, we consider the
three phenomenological motivations of our study. We will see that
the controversies can be resolved all together by simply assuming
the existence of a large viscosity, which slows down all the
corresponding processes. We derive from the structure of the
landscape the origin of the viscosity in section 3. And in section
4, we try to illustrate our cosmic fuzzball scenario by
reconstructing from the landscape data the accelerating geometry,
which is subsequently compared to the static coordinate metric of dS
space. Section 5 includes some discussions of future directions.
Recent discussions of the quantum landscape can be found in \kt.

\newsec{Phenomenological issues}

The phenomenological motivation of our study is threefold, the
cosmological constant problem, moduli stabilization and the initial
condition for inflation.

Direct observation of the acceleration of the universe came from the
supernova project \sn, and many puzzles were brought out by their
observations. The first puzzle is that the universe is accelerating.
The universal attractive force imposed by the matter and radiation
predicts a deceleration of the universe. A possible source of the
repulsive force is the zero point energy of quantum fields. Assuming
this, there comes the second puzzle that the acceleration is much
slower than any estimation employing quantum field theory. This is
the widely realized cosmological constant problem, which is often
asked as why the cosmological constant is so small. The deviation of
the theoretical estimation from the experimental results is usually
illustrated by the ratio of the energy density
\eqn\cosmp{{{\rho_V^{the}}\over {\rho_V^{exp}}}\approx 10^{-120}.}
Here we propose that the extremely small cosmological constant, or
the extremely slow acceleration of the probe objects, such as the
supernova, is a result of an effective extremely large viscosity
experienced by these objects.

In a simple Newtonian model, consider that the cosmological constant
provide a repulsive force and the viscosity a friction term
\eqn\nt{m\dot{v}=f_{\Lambda}-g(v),} where $g(v)$ is some function of
the velocity $v$. If $g(v)$ is linear in $v$ \eqn\vvv{g(v)=bv,} the
acceleration is an exponentially decaying function
\eqn\livel{\dot{v}=e^{-{b\over m}(t-t_0)}.} What's important is that
the acceleration is now controlled by friction rather than by the
cosmological constant term. And the question why the acceleration
$\dot{v}$ is so small is converted to the question why the friction
$b$ is so large.

%When the viscosity is sufficiently large, thus $b/m$ large enough,
%the behavior of $\dot{v}$ in \livel\ resembles that of a step
%function. That is, for $t<t_o$, $\dot{v}$ is divergently large,
%while for $t>t_0$, $\dot{v}$ is set to nearly zero.
%Phenomenologically, the $t<t_0$ portion corresponds to the epoch of
%inflation, where all harmful defects are exponentially diluted,
%while the $t<t_0$ portion corresponds to the present epoch of slow
%acceleration, where dark energy dominates. This picture is obviously
%very crude. Actually between these two acceleration phases, there
%exist phases dominated by matter and radiation. Nevertheless it is
%interesting to embed inflation and dark energy into a single
%framework.

More precisely, consider a universe with matter whose energy density
is $\rho={\rho_0\over{a^3}}$, and cosmological constant $\Lambda$,
the Friedmann equation reads \eqn\fri{({\dot{a}\over a})^2={{8\pi
G_N}\over3}{\rho_0\over a^3}-{k\over a^2}+{\Lambda\over3},} where
$a$ is the scale factor and $k$ the curvature in the
Robertson-Walker metric
\eqn\rwm{ds^2=-dt^2+a^2(t){{d\vec{x}^2}\over{(1+{1\over4}k\vec{x}^2)}^2}.}
Mathematically the Friedmann equation can be looked upon as the
equation of motion for a particle moving in one dimension subjected
to the potential \eqn\fppp{V(a)=-{{4\pi G_N}\over3}{\rho_0\over
a}+{k\over 2}-{\Lambda\over6}a^2.} With large $a$, the cosmological
constant term dominates and the expansion of the universe
accelerates. When there is extremely large viscosity, the motion in
the effective potential is greatly slowed down, and $\dot{a}\over a$
becomes very small, which, for an observer who neglects the
viscosity, will be ascribed to an extremely small cosmological
constant. We propose that this is the origin of the apparent
smallness of the cosmological constant.

In the study of string phenomenology, the stabilization of moduli
\gkp\ \kklt\ is a requisite to get a realistic model, which has
nevertheless proven to be quite difficult to achieve. However in the
present approach, this is not necessary, since all long distance
dynamics are slowed down.

The problem of moduli stabilization can be illustrated as follows.
Compactify type IIB string theory to 4 dimensions, and consider the
typical effective potential \lin \eqn\mpo{V(\phi,\rho,\psi)\sim
e^{\sqrt2\phi-\sqrt6\rho}\tilde{V}(\psi),} where $\phi$ and $\rho$
are normalized dilaton field and volume of internal space, while
$\psi$ represents other fields. The dilaton will rapidly go to
$-\infty$, and the volume modulus to $+\infty$ due to the
exponential factor $e^{\sqrt2\phi-\sqrt6\rho}$. So the 4 dimensional
space will rapidly decompactify to 10 dimensions.

However if the motion in potential \mpo\ is subject to large
viscosity, the decompactification time scale will be much prolonged
and this process will not be detrimental to model building. Thus
moduli stablilization becomes unnecessary. In fact, the dS vacua
with all moduli stabilized as constructed in \kklt\ are also
metastable and there it was only required that their lifetimes are
long enough. Without need of moduli stabilization, we hope that
building string phenomenological models can be much easier.

Here a problem of our mechanism arises. If the universe starts in a
state with dimension more than four, it seems that it can stay high
dimensional for a long time. We have to assume that the smallness of
spacetime dimension, as well as the smallness of the rank of the
gauge group, can be explained by some mechanism, in this or some
other framework.

Another motivation of moduli stabilization is the hierarchy problem
\gkp. Randall-Sundrum's idea of generating hierarchy via redshift
\rs\ was realized by Verlinde in string theory \verl, where D3
branes are placed on a compact manifold, and a warp factor is
generated \eqn\warp{e^{-4A}\approx{{4\pi g_sN}\over{\tilde{r}^4}},}
with $N$ number of branes, $g_s$ the string coupling, $\tilde{r}$
distance from the D3 branes. $\tilde{r}$ has no potential to fix it,
and to get a stable hierarchy, fluxes are added \gkp. The resulted
hierarchy of energy scale is controlled by the fluxes
\eqn\warpf{e^A\sim \exp(-2\pi K/3Mg_s),} with $K, M$ the fluxes.
When time evolution of $\tilde{r}$ is sufficiently slowed down by
large viscosity, one can get a stable hierarchy without fluxes.

With long distance dynamics slowed down, the slow roll inflation
also becomes natural. Consider a nearly homogeneous scalar field
$\psi$ coupled to gravity with action \eqn\ipo{S=\int d^4x
\sqrt{-g}[\half g^{\mu\nu}\p_{\mu}\psi\p_{\nu}\psi-V(\psi)].} The
energy density and pressure
are\eqn\rhp{\eqalign{\rho_{\psi}=\half\dot{\psi}^2+V(\psi), \cr
p_{\psi}=\half\dot{\psi}^2-V(\psi).}} If the scalar field motion is
slowed down due to large viscosity, making $\dot{\psi}^2\ll
V(\psi)$, or $\rho_{\psi}\simeq-p_{\psi}$, the equation of state of
$\psi$ will approximate that of the cosmological constant. Thus
inflation occurs. The stringent constraint on the form of the
potential $V(\psi)$ is much loosed this way.

And the above mechanism applies equally well for the emergence of
dark energy. The sole difference between the epoch of inflation and
the epoch of current slow acceleration dominated by dark energy is
the energy scale of the effective cosmological constant. Thus our
mechanism can explain

1), why there exists dark energy,

 2), why dark energy is so small.

But it is more subtle to understand inflation. The above mechanism
provides the $w={P\over\rho}=-1$ matter that can drive inflation,
but if the evolution of the scale factor $a$ is also similarly
slowed down as in the case of dark energy, there will be no
sufficiently fast process like inflation. One way to solve this
problem is to assume that the scale factor behaves differently as
the matter degrees of freedom. In the inflationary phase, the scale
factor responds to the vacuum as if it is a perfect fluid, with very
small viscosity, while the other degrees of freedom respond as if
the vacuum is a glass. And the scale factor part also enters the
glass phase when inflation ends and the universe goes through the
epoch dominated by matter and radiation and gradually enters the
current slow acceleration phase.

Actually the community of high temperature superconductors have
witnessed the separation of degrees of freedom, the spin-charge
separation, which says that the spin degrees of freedom in a medium
can respond differently as the charge degrees of freedom in the same
medium. In the study of quantum decoherence of the universe \kiefer,
the metric also behaves differently as the other degrees of freedom,
such as the matter fields. The former plays the role of the
microscopic object and the latter the role of the environment, whose
presence results in the loss of quantum coherence. Quantum
decoherence is also a dissipative dynamics, similar to the effect of
viscosity.

A crucial requisite of the success of the above explanations of the
three problems is that short distance dynamics remains the same,
described by flat space quantum field theory plus general relativity
with asymptotic Minkowski background. Here by short we mean
non-cosmological, for example the physics on the earth, or in the
solar system, where numerous experiments have been done to check the
validity of the existing theories. We only modify the theory at the
cosmological scale. However it is generally not easy to modify
general relativity only at large distance. In the following we will
show that in our framework, this is naturally achieved due to the
separation of time scales. In the study of glass transition, the
separation of time scales was interpreted by suggesting that
molecular motions consist of anharmonic vibrations about deep
potential energy minima and of infrequent visitations of different
such minima \gol, which is the prototype of the landscape scenario
in studying the glass state.

\subsec{A model for scalar field driven inflation}

Now consider a viscous scalar field coupled to gravity. Assume that
the potential energy density has the simple form
$V(\psi)={1\over2}m^2\psi^2$. The energy momentum tensor in the
presence of gravitation has the generally covariant form
\eqn\covstr{T^{\mu\nu}=pg^{\mu\nu}+(p+\rho)U^{\mu}U^{\nu}-\eta
H^{\mu\rho}H^{\nu\sigma}W_{\rho\sigma},} where
\eqn\wh{W_{\mu\nu}\equiv
U_{\mu;\nu}+U_{\nu;\mu}-{2\over3}g_{\mu\nu}U^{\gamma}_{;\gamma}}and
\eqn\hhhh{H_{\mu\nu}\equiv g_{\mu\nu}+U_{\mu}U_{\nu}.}Here we follow
the notation of Weinberg \wein.

The dissipative terms vanish for a Robertson-Walker metric, so the
Friedmann equation remains the same as the case without viscosity
\eqn\friv{H^2+{k\over a^2}={1\over6}(\dot{\psi}^2+m^2\psi^2).} The
equation of motion for the scalar field is modified by adding a
viscous term besides the cosmological damping term
\eqn\vissca{\ddot{\psi}+3H\dot{\psi}+2\eta\dot{\psi}=-m^2\psi,}
where we set $M_{pl}^2=8\pi G_N=1$. Since the viscosity $\eta$ is
expected to be large, the scalar field moves slowly while the scale
of the universe grows rapidly. Subsequently, one has $\ddot{\psi}\ll
3H\dot{\psi}, H^2\gg {k\over a^2}, \dot{\psi}^2\ll m^2\psi^2$, and
the above two equations simplifies to \eqn\hsim{H={\dot{a}\over
a}={{m\psi}\over{\sqrt
6}},}\eqn\psisim{3H\dot{\psi}+2\eta\dot{\psi}=-m^2\psi.}Combining
these two equations, one gets the equation for $\psi$ to be
\eqn\ppsi{{\sqrt6\over2}m\psi\dot{\psi}+2\eta\dot{\psi}+m^2\psi=0.}And
the scale factor turns out to be \eqn\aaapsi{a=a_0
\exp({1\over4}\psi^2+{\sqrt6\over3}{\eta\over m}\psi).}The first
term in the exponential factor comes from the cosmological damping
term $3H\dot{\psi}$, and the second term from the contribution of
the viscosity. Here we note that, compared to chaotic inflation, one
does not need to assume the scalar field $\psi$ to be initially
large. Due to large $\eta$, even small changes in $\psi$ will result
in exponentially large expansion of the scale factor $a$. To get a
reasonably large number of e-foldings, eg. $N\sim 60$, assuming the
change in the value of the scalar field to be as small as
$\psi\sim1$, the viscosity is required to be no less than
$\eta/m\sim 74$, which is easily satisfied. Thus the constraint on
the initial conditions for inflation is now more loosed.

We note also that our mechanism shares the same merit as the chaotic
inflation model that it is quite generic in the sense that it does
not depend on the particular form of the potential.

\subsec{A model for the small cosmological constant}

Consider the universe driven mainly by a large cosmological
constant, which comes for example from a slowly rolling scalar
field. As mentioned above, we assume that after inflation, a new
phase emerges where the scale factor is also subject to a large
viscosity, so the Einstein equation is modified. The spatial
component now has the form \eqn\addot{{\ddot{a}\over
a}+2\eta{\dot{a}\over a}=-{1\over6}(\rho+3p)={\Lambda\over3}.}

The Friedmann equation is also changed, but for an observer who
disregards the viscosity will define the effective cosmological
constant through \eqn\effcc{({\dot{a}\over
a})^2={\Lambda_{eff}\over3}.} So the effective cosmological constant
is related to the bare cosmological constant via
\eqn\ccecc{{\Lambda_{eff}\over3}+2\eta\sqrt{\Lambda_{eff}\over3}={\Lambda\over3},}
or
\eqn\eccecc{\Lambda_{eff}=3(\sqrt{\eta^2+{\Lambda\over3}}-\eta)^2.}
In the next section, we will show that an exponentially large
viscosity can be achieved, so the above equation can be expanded to
be \eqn\eecc{\Lambda_{eff}\simeq{\Lambda^2\over{12\eta^2}}.} Assume
that the bare cosmological constant is of the Planck scale
$\Lambda\sim M_{pl}^2$, and the viscosity of the form
\eqn\ccef{\eta=M_{pl}e^{N_{cc}},} to get the large hierachy \cosmp,
the number of e-foldings for the cosmological constant is required
to satisfy \eqn\ccefold{N_{cc}\geq68.}

What's important here is that the form of the geometry remains the
same as the case with a bare cosmological constant without
viscosity, the Robertson-Walker metric with
$a(t)=a_0\exp(\sqrt{\Lambda_{eff}/3}t)$, so in leading order, our
mechanism has the same phenomenology as the model with a small
cosmological constant, thus consistent with current observations.
However the naturalness of the smallness of the cosmological
constant can be understood in this framework.

\subsec{A model for the slowly-rolling moduli}

Consider for example the volume modulus $\rho$ with potential
$V(\rho)\sim e^{-\sqrt6\rho}$. In the following, We disregard the
influence of the other moduli and also ignore the numerical factors
of order 1. So the equation of motion for $\rho$ reads
\eqn\eomrho{\ddot{\rho}+\eta\dot{\rho}+e^{-\sqrt6\rho}=0.} With
large viscosity, one has $\ddot{\rho}\ll\eta\dot{\rho}$, and the
time cost to move from $\rho_1$ to $\rho_2$ reads \eqn\modttt{\Delta
t=\eta (e^{\sqrt6\rho_1}-e^{\sqrt6\rho_2}).} Restoring $M_{pl}$, the
time cost for $\rho$ to roll from $V(\rho)=1$ to $V(\rho)=\half$ is
\eqn\timem{\Delta t\sim{\eta\over M_{pl}^2}.} To achieve a life time
of the order the cosmological time scale $\sim 10^{10}$ years, the
number of e-foldings needs to satisfy
\eqn\modecc{N_{cc}\geq140.}Other moduli can be considered the same
way as above, and one can get the corresponding bounds for $N_{cc}$,
which are of the same order as \modecc.

\newsec{Slowing down in the landscape}

In the following we focus on the question how the existence of the
landscape leads to the slow dynamics and separation of time scales.

Some features of the string landscape were recently explored in
\cdgkl. BPS domain walls were found where the potential has more
than one critical points, with and without barrier between them. The
critical points may be AdS, dS or Minkowski vacuum, and they may be
minima, maxima or saddle points. The importance of the decay of
metastable dS space was emphasized in \cdgkl. Besides the reversible
transition between different dS vacua, it was noticed that the decay
of dS space into Minkowski space or AdS space via bubble production
is irreversible. And such irreversible channels of vacuum decay were
interpreted in their framework as suggesting the existence of sinks
for flow of probability in the string landscape.

Here we further assume some general characteristics of the string
landscape parallel to the previous work \gol\ in supercooled glass
forming liqiuds. First, we assume that there are a large number of
minima in the landscape, and these minima may have varying depths.
Various such vacua were found since the work of \gkp\ \kklt\ in the
study of flux compactification \dk.

Second, our universe is at or near a minimum of the potential. The
excitations about the minimum are described by quantum field theory
and also the Einstein-Hilbert action.

Third, quantum fluctuations make possible transitions between
different minima. In \kklt\ two transition mechanisms were
discussed, the Coleman-de Luccia(CdL) bubble \cdl\ and the
Hawking-Moss(HM) instanton \hm. The former works when the potential
energy barrier is narrow relative to its height. Otherwise the
latter dominates the transition. Thermal production of membranes can
also lead to transition between vacua with different cosmological
constants \bt\ \bp\ \ghtw. Details of the transitions are not that
important, but their time scales will have effect.

The fourth assumption is that quantum fluctuations are not very
severe, or stated more precisely, average energy of zero point
fluctuations $\hbar\bar{w}$ is smaller than or at most comparable to
height of the potential energy barrier $\Delta$
\eqn\ass{\hbar\bar{w}\leq\Delta ,} so that the properties of the
system will be dominated by long residences near local minima of the
potential energy with rare transition events between the minima.

The viscosity is the time integral of the stress correlation
function. In a glass state, the local stress fluctuations do not
decorrelate in time, and this leads to large viscosity. In the
notion of landscape, this comes from the fact that the system visits
only a finite number of configurations with appreciable probability,
which is characteristic of broken ergodicity \gger. With
sufficiently large quantum fluctuations, all configurations in the
landscape can be explored with appreciable probability. In this
case, the stress correlation function goes to zero in the long time
limit, and the system is in a perfect fluid state where ergodicity
is restored.

The slow dynamics in a glassy state can be qualitatively understood
as a result of the high dimensionality of the landscape \kurc. As
mentioned above \cdgkl, there are many critical points in the
landscape, where gradient of the energy vanishes. A critical point
is characterized by the number of negative eigenvalues of the energy
Hessian \eqn\hes{H_{ij}={{\p^2E}\over{\p s_i\p s_j}},} where $s_i$
labels a configuration in the landscape. We call this number the
index $I$ of the point. The minima have $I=0$, and the maxima $I=N$,
with $N$ the dimension of the landscape. And we assume generally
that there are critical points of every index.

The landscape can be tiled into different parts with respect to
these critical points. The set of points that will flow through
gradient descent to a minimum are grouped into a basin of
attraction. Points on the $N-1$ dimensional border $\p_1$ of a basin
will be stuck on the border. The trajectories of these points will
end in minima on $\p_1$, which have index $I=1$, the saddle points.
Thus $\p_1$ is itself tiled into basins of attraction with borders
of dimension $N-2$, denoted as $\p_2$. And the process can be
reiterated to get a series of borders $\p_1, \p_2, \p_3, ..., \p_J$.

The key point is that in a high dimensional landscape, a random
starting point in a basin will be very close to a border. The
distance between two configurations can be defined as
\eqn\dist{D(a,b)={1\over N}\sum^N_{i=1}(s^a_i-s^b_i)^2,} and most of
the volume of a basin is contained within a distance of
$D\simeq{1\over N}$ from its border $\p_1$ \kurc. And similar
arguments work for $\p_2, ..., \p_J$. While systems starting on
$\p_J$ will be stuck at a critical point of index $J$, a randomly
chosen system, which generally starts near $\p_J$, will be trapped
in the neighborhood of a critical point with index $I=J$, and the
dynamics of the system are slowed down.

The slowing down mechanism can also be illustrated by considering a
gaussian model, where the energy distribution of the minima in the
landscape has an explicit functional form
\eqn\edf{\Omega_N(E)=\exp(\alpha
N)\exp[-{{(E-E_0)^2}\over{\epsilon^2}}],} where $N$ is the dimension
of the landscape, $E_0$ some reference energy, $\alpha$ and
$\epsilon$ two parameters. Here we follow the notation of \rsz. The
configuration entropy defined as $\Sigma(e)=\ln \Omega_N(E)$, now
reads
\eqn\cfe{\Sigma(e)=N[\alpha-{{(e-e_0)^2}\over{\bar{\epsilon}}}],}
with $\bar{\epsilon}=\epsilon/{\sqrt N}$ and $e=E/N$.

A crucial input is the Adam-Gibbs equation \ag\ which establishes a
relation between the viscosity and the configuration entropy. In our
case where thermal fluctuations are replaced by quantum
fluctuations, the equation has the form
\eqn\qage{\eta(\bar{w})=\eta_\infty
\exp({E_1\over{\hbar\bar{w}\Sigma(\bar{w})}}),} with $\eta_\infty$
the infinite fluctuation limit of the viscosity and $E_1$ is related
to the energy barrier. The viscosity diverges at the zero points of
the configuration entropy, where
\eqn\crie{e_*=e_0-\bar{\epsilon}\sqrt \alpha.}This is qualitatively
easy to understand. When the configuration entropy goes to zero,
there are less and less available configurations for the system to
explore, and the dynamics slows down. In this simplified model, we
can make precise the definition of the averaged zero point energy as
\eqn\gzpe{{1\over{\hbar\bar{w}}} ={{d\Sigma(e)/N}\over{de}},} in
analogy with the definition of temperature in statistical mechanics
as $T^{-1}=dS/dE$. At the zero entropy points, the zero point energy
is \eqn\zpe{\hbar w_*={\bar{\epsilon}\over{2\sqrt\alpha}}.} Thus the
viscosity reads
\eqn\visag{\eta(\bar{w})=\eta_\infty\exp(\gamma{w_*\over{\bar{w}-w_*}}{\bar{w}\over{\bar{w}+w_*}}),}
where $\gamma=E_1/(\alpha N\hbar w_*)$. So when the magnitude of
quantum fluctuations is at or near the critical value, the system
will have divergently large viscosity. We call this state a quantum
glass state to distinguish it from the glass states driven by
thermal fluctuations.

The universe is expected to relax to a ground state, and the
relaxation time has also the Adam-Gibbs form \ag
\eqn\reag{\tau(\bar{w})=\tau_\infty
\exp({E_1\over{\hbar\bar{w}\Sigma(\bar{w})}}).} Similar to
viscosity, it reads
\eqn\relag{\tau(\bar{w})=\tau_\infty\exp(\gamma{w_*\over{\bar{w}-w_*}}
{\bar{w}\over{\bar{w}+w_*}}),} in the gaussian model. We see that
that the relaxation time can be very large in contrast to an
intuitive estimation.

Separation of time scales is in fact closely related to the
largeness of the viscosity. The lifetimes of elementary excitations
are essentially the relaxation times associated with shear and bulk
relaxation processes \zw. For a relativistic field theory, the shear
viscosity $\eta$ has the form \hst
\eqn\visc{\eta=\lim_{w\rightarrow0}{1\over{2w}}\int
dtd\vec{x}e^{iwt}<T_{xy}(t,\vec{x})T_{xy}(0,0)>,} where $T_{xy}$ is
the $xy$ component of the stress tensor, and
\eqn\sss{<T_{-k}^{xy}(0)T_k^{xy}(0)>=V^{-1}G_{\infty},} with
$G_{\infty}$ the high-frequency shear modulus. The long-wave limit
of the lifetime of the transverse modes is connected with viscosity
via \zw \eqn\lif{\tau_1={\eta\over G_\infty}.} And the
long-wavelength limit of the longitudinal lifetime is similarly
connected with the bulk viscosity. So in a glassy state with large
viscosity, the excitations with frequencies low relative to $\tau_1$
will be highly damped, and contributions to macroscopic properties
coming from higher frequency excitations are well separated from
those from lower frequencies \gol.

From the landscape point of view, the short time dynamics is related
to the intra-basin motion around a local potential energy minimum,
while long time dynamics is related to inter-basin transitions. With
decreasing quantum fluctuations, deeper energy areas will be
explored but with lower degeneracy, thus decreasing configuration
entropy, which signals the ordering of the system in the landscape.

\newsec{Emergence of the geometry}

We see from the Adam-Gibbs relation above that dynamics in the glass
state are all slowed down by an exponential factor
$\exp({E_1\over{\hbar\bar{w}\Sigma(\bar{w})}})$, and this effect is
in fact geometric, in the sense that it can be capsulated by
multiplying time by a redshift factor \eqn\gred{dt\rightarrow
\exp(-{E_1\over{\hbar\bar{w}\Sigma(\bar{w})}})dt.} When the
configuration entropy $\Sigma(\bar{w})$ vanishes, the redshift
factor goes to zero, and time is frozen. This is the defining
property of the event horizon in gravitational systems.

 Here we view
the spacetime geometry as an emergent concept, and try to
reconstruct the metric from the magnitude of the prolonging of time
scales. The location of the horizon is where the glass transition
happens. The appearance of a horizon signals the existence of
thermal effects, and Hawking temperature can be defined as
\eqn\hawk{k_BT_H=\hbar \bar{w}.} So the above derivation of the
viscosity \visag\ and relaxation time \reag\ in the Gaussian model
completely parallels those in the supercooled glass forming liquids.
The Hawking temperature at the horizon is determined by the critical
value of the quantum fluctuations, and subsequently the energy
distribution of the minima in the landscape
\eqn\hawh{k_BT_H^{*}=\hbar w_*={\bar{\epsilon}\over{2\sqrt\alpha}}.}

Up to this point, we have established two results that glass
transition happens at the horizon scale and local physics remains
the same. Obviously they are viewed from the perspective of a static
observer. And the specific characteristic of cosmic horizon is also
well accounted for, where every static observer is attached with a
horizon.

We will compare the static coordinate metric of dS space
\eqn\dsst{ds^2=-(1-{r\over l})dt^2+({1-{r\over
l}})^{-1}dr^2+r^2(d\theta^2+\sin^2\theta d\phi^2)} with the above
result. The cosmic horizon lies at $r=l$. In dS space, the Hawking
temperature detected by a geodesic observer is connected with the dS
radius $l$ as \eqn\dstem{T_{dS}={1\over{2\pi l}},} while the
cosmological constant is \eqn\ccds{\Lambda={3\over l^2}.}

The redshift factor in the Gaussian model is \eqn\rec{f(T_H)=
\exp(-\gamma{{T_HT_H^*}\over{T_H^2-{T_H^*}^2}}),} and we identify
$T_H^*=T_{dS}={1\over{2\pi l}}$. To get an energy hierarchy of order
$10^{-40}$, as in the case of the cosmological constant problem
represented in equation \cosmp, the tuning is much looser. The
temperature is required to be in the range
\eqn\tune{{{T_H-T_H^*}\over T_H^*}\sim {\cal O}(10^{-2}\gamma).}

Assuming spherical symmetry, a metric can be reconstructed from the
redshift factor with the form
\eqn\remet{ds^2=-f^2(T_H)dt^2+f^{-2}(T_H)dr^2+r^2(d\theta^2+\sin^2\theta
d\phi^2).} Hawking temperature $T_H$ varies with radius $r$
\eqn\ttr{T_H=T_H(r).} The radial direction can be viewed as
holographically emergent from the hydrodynamics of the landscape,
since every radial coordinate $r$ can be identified by the
corresponding Hawking temperature $T_H$ through \ttr.

This metric is quantitatively very different from the dS space
metric \dsst, which we view as the result of the fact that the
gaussian model is oversimplified, but qualitatively they have many
similarities. The limit $T_H\rightarrow T_H^*$ corresponds to
$r\rightarrow l$, stating that a horizon emerges at the glass
transition point. The limit $T_H\rightarrow\infty$ corresponds to
$r\rightarrow0$, which means that physics which happens at the scale
smaller than the horizon scale is in the high temperature phase,
with small viscosity and small redshift. As a first order
approximation, we identify the redshift factor calculated from the
gaussian model with that of dS space between these two limits
\eqn\fth{ \exp(-\gamma{{T_HT_H^*}\over{T_H^2-{T_H^*}^2}})\simeq
1-{r\over l}, ~~~~~ 0<r<l.}

It may seem strange that the $r\rightarrow0$ limit, which is locally
Minkowski, has infinite temperature. We note that the locally
Minkowski nature of the emergent spacetime is a result of the random
superposition of the configurations in the landscape, and the
infinite temperature signals the total randomness of the
superposition. The geometry is not treated as a fixed background as
in quantum field theory in curved space. This can also be roughly
understood as that our universe is more like the part of the
spacetime inside the black hole horizon, where temperature can
certainly be very high. And the correspondence of the temperature
$T_H$ with the radial coordinate $r$ is also a UV/UV connection in
accordance with the result found by probing deep inside the black
hole horizon \hliu, in contrast to the familiar UV/IR relation in
AdS/CFT \suwi, where long distance in the bulk corresponds to high
energy in the boundary.

More notable is the difference between the two metrics. The function
$f(T_H)$ is not continuous at the point $T_H=T_H^*$, and there is an
infinite jump from $+\infty$ to $0$. The redshift factor $1-{r\over
l}$ in the metric \dsst\ changes continuously when r goes across the
point $r=l$. We interpret this as a sign that the metric \dsst\
breaks down outside the horizon where $r>l$. Since there is a phase
transition at the horizon, the metric inside the horizon can not
directly be continuously extended to the region outside the horizon,
where a new glass phase appears.

\newsec{Discussions}

In this note, we studied the collective dynamics of the landscape,
and find that large scale dynamics slows down while short distance
dynamics remains the same as before. The warping generated in the
glass state provides a natural mechanism for the accelerating
universe to emerge from the inherent structure of the landscape. The
slowing down effect can be used to understand the cosmological
constant problem, the naturalness of the slow roll inflation in the
landscape and the overshot problem in string compactification.

We hope to find scenarios other than the anthropic principle to
understand the nature of our universe. And obviously the scenario
outlined in this note is rude and in its infancy. Many open
questions are left behind. First we hope we can have better control
of the landscape and reconstruct a more realistic geometry of the
universe. One needs to go beyond the simple gaussian model, and
hopping effects among the minima may also play an important role.

The second question is whether other quantities of our universe, for
example the dimension of spacetime and rank of the gauge group, can
also be understood in this scenario. Maybe we can reparameterize the
question as why these two quantities are so small. In fact the
cosmological constant is also a quantized number as these two
quantities. To begin with, we may first make these quantities
continuous.

Another approach to understand the glass transition is based on the
concept of frustration. A simple example is the Ising spin model on
a triangular lattice with antiferromagnetic interactions between
nearest neighbor spins, where the spins can not minimize the energy
of the system by merely minimizing all local interactions due to the
tiling of whole space. Frustration leads to two long time scales,
which can subsequently help to understand the hierarchy problem
\kivel. In the absence of frustration, there is the correlation
length $\xi_0$ of the fluctuations. When frustration is turned on,
the locally favored structure can not tile the whole space, and
domain walls are formed. The typical size $R_D$ of the frustration
induced domain walls is another long length scale, and it is
generally much larger than the correlation length $\xi_0$. The
redshift factor has the form \eqn\lnred{\ln
f(T)\propto({R_D\over\xi_0})^2{T^*\over T},} and an exponentially
large hierarchy is generated.

We speculate here that the competition between the local quantum
field theory and the holographic constraint of the number of degrees
of freedom results in a frustration. This helps us to understand the
difficulty of a longstanding problem of AdS/CFT, that is how to
extract bulk local physics from the dual field theory. From the
frustration point of view, the two are both fundamental, not that
one can be derived from the other, and actually they are competing.
It will also be interesting if we can explore the nature of
cosmology from the view of frustration.

\bigskip

Acknowledgments.

We thank D. W. Pang for discussions that initiated this project. And
we also thank M. Li and Q. G. Huang for valuable comments.

\listrefs
\end